\documentclass[12pt,twoside,english,authoryear]{elsarticle}
\usepackage[T1]{fontenc}
\usepackage[utf8x]{inputenc}
\usepackage{color}
\usepackage{babel}
\usepackage{amsthm}
\usepackage{amsmath}
\usepackage{esint}
\usepackage[unicode=true,
 bookmarks=true,bookmarksnumbered=true,bookmarksopen=true,bookmarksopenlevel=99,
 breaklinks=false,pdfborder={0 0 1},backref=section,colorlinks=true]
 {hyperref}
\hypersetup{pdftitle={The LyX Tutorial},
 pdfauthor={LyX Team},
 pdfsubject={LyX-documentation Tutorial},
 pdfkeywords={LyX, documentation},
 linkcolor=black, citecolor=black, urlcolor=blue, filecolor=blue,pdfpagelayout=OneColumn, pdfnewwindow=true, pdfstartview=XYZ, plainpages=false}
\usepackage{breakurl}

\makeatletter
\numberwithin{equation}{section}
\numberwithin{figure}{section}

\usepackage{ifpdf} 
\ifpdf 

 \IfFileExists{lmodern.sty}{\usepackage{lmodern}}{}

\fi 

\let\myTOC\tableofcontents
\renewcommand\tableofcontents{%
  \frontmatter
  \pdfbookmark[1]{\contentsname}{}
  \myTOC
  \mainmatter }

\usepackage{breqn}
\gdef\wrap@breqn@environ#1#2{
    \expandafter\let\csname breqn@oldbegin@#1\expandafter\endcsname\csname #1\endcsname
    \expandafter\let\csname breqn@oldend@#1\expandafter\endcsname\csname end#1\endcsname
    \expandafter\gdef\csname breqn@begin@#1\endcsname{%
        \expandafter\let\csname #1\expandafter\endcsname\csname breqn@oldbegin@#1\endcsname%
        \begin{#2}%
    }
    \expandafter\gdef\csname breqn@end@#1\endcsname{%
        \expandafter\let\csname end#1\expandafter\endcsname\csname breqn@oldend@#1\endcsname%
        \end{#2}%
        \expandafter\let\csname #1\expandafter\endcsname\csname breqn@begin@#1\endcsname%
        \expandafter\let\csname end#1\expandafter\endcsname\csname breqn@end@#1\endcsname%
    }
    \expandafter\let\csname #1\expandafter\endcsname\csname breqn@begin@#1\endcsname
    \expandafter\let\csname end#1\expandafter\endcsname\csname breqn@end@#1\endcsname
}
\wrap@breqn@environ{equation}{dmath}
\wrap@breqn@environ{equation*}{dmath*}
\usepackage{cancel}
\makeatletter
\def\ps@pprintTitle{%
  \let\@oddhead\@empty
  \let\@evenhead\@empty
  \def\@oddfoot{\reset@font\hfil\thepage\hfil}
  \let\@evenfoot\@oddfoot
}
\makeatother

\makeatother

\begin{document}

\title{Multiplying decomposition of stress/strain, constitutive/compliance
relations, and strain energy}

\author{Lee, HyunSuk\fnref{lee}}

\author{Jinkyu Kim\fnref{kim}}

\fntext[lee]{Ph.D. student, Department of Civil, Structural and Environmental
Engineering, Ketter Hall, University at Buffalo, Buffalo, NY, 14260,
USA. Email: hl48 @ buffalo.edu }

\fntext[kim]{Research Professor, School of Civil, Environmental and Architectural
Engineering, Korea University, Anam-dong 5-ga 1, Seongbuk-goo 136-713,
Korea (corresponding author). Tel: 82-2-3290-3833, Fax: 82-2-921-2439,
Email: jk295@korea.ac.kr }
\begin{abstract}
To account for phenomenological theories and a set of invariants,
stress and strain are usually decomposed into a pair of pressure and
deviatoric stress and a pair of volumetric strain and deviatoric strain.
However, the conventional decomposition method only focuses on individual
stress and strain, so that cannot be directly applied to either formulation
in Finite Element Method (FEM) or Boundary Element Method (BEM). In
this paper, a simpler, more general, and widely applicable decomposition
is suggested. A new decomposition method adopts multiplying decomposition
tensors or matrices to not only stress and strain but also constitutive
and compliance relation. With this, we also show its practical usage
on FEM and BEM in terms of tensors and matrices. \end{abstract}
\begin{keyword}
Pressure \sep hydrostatic pressure \sep Deviatoric stress \sep
Volumetric strain \sep Mean strain \sep Deviatoric strain \sep
Stress decomposition \sep Strain decomposition \sep Constitutive
decomposition \sep Compliance decomposition \sep Multiplication
decomposition \sep Decomposition multiplier
\end{keyword}
\maketitle

\section{Introduction}

In many references (see \citet{fung1965foundations,fung2001classical,gurtin1981introduction,richards2000principles,bower2009applied}),
stress and strain decomposition are usually used for a set of invariants.
From a hydrostatic stress tensor (or volumetric stress tensor) $p$,
the 1st stress invariant $I_{1}$ is given by 

\begin{equation}
I_{1}=3\; p\label{eq:1_1}
\end{equation}

Also, from a deviatoric stress tensor $s_{ij}$, the second and third
deviatoric stress invariants are given by

\begin{eqnarray}
J_{2} & = & \frac{1}{2}\; s_{ij}\; s_{ij}\label{eq:1_2}\\
J_{3} & = & \frac{1}{3}\; s_{ij}\; s_{jk}\; s_{ki}\label{eq:1_3}
\end{eqnarray}

In \ref{eq:1_1}-\ref{eq:1_3}, a hydrostatic stress tensor $p$ and
a deviatoric stress tensor $s_{ij}$ result from the decomposition
of stress tensor $\sigma_{ij}$:

\begin{eqnarray}
\sigma_{ij} & = & s_{ij}+p\;\delta_{ij}\label{eq:1_4}\\
p & = & \frac{\sigma_{kk}}{3}\label{eq:1_5}
\end{eqnarray}

Similarly, from a volumetric strain tensor $\epsilon_{M}$ ($\epsilon_{M}=\epsilon_{kk}/3$)
and a deviatoric strain tensor $\epsilon_{ij}^{'}$ ($\epsilon_{ij}^{'}=\epsilon_{ij}-\epsilon_{M}\;\delta_{ij}$
), we have strain invariants $I_{1}^{\epsilon}$, $J_{2}^{\epsilon}$,
and $J_{3}^{\epsilon}$ as

\begin{eqnarray}
I_{1}^{\epsilon} & = & 3\;\epsilon_{M}\label{eq:1_6}\\
J_{2}^{\epsilon} & = & \frac{1}{2}\;\epsilon_{ij}^{'}\;\epsilon_{ji}^{'}\label{eq:1_7}\\
J_{3}^{\epsilon} & = & \frac{1}{3}\;\epsilon_{ij}^{'}\;\epsilon_{jk}^{'}\;\epsilon_{ki}^{'}\label{eq:1_8}
\end{eqnarray}

The other general use of stress and strain decomposition can be found
in phenomenological theories (e.g., \citet{NEWhoulsby2002rate,perzyna1966fundamental,NEWdunne2005introduction,NEWsimo1998computational,NEW_lubliner1990plasticity,NEW_Leighl}).
For example, von Mises yield function $f$ that describes elasto-plasticity
is written as

\begin{eqnarray}
f & = & J_{2}-k^{2}\label{eq:1_9}
\end{eqnarray}

where, $k^{2}=\frac{1}{3}\;\sigma_{Y}^{2}$. Also, the flow potential
$\varphi$ in Perzyna formulation that describes rate-dependent plasticity
is written as

\begin{eqnarray}
\varphi & = & \frac{1}{2\;\eta}\;\left\langle \sqrt{J_{2}}-k\right\rangle ^{2}\label{eq:1_10}
\end{eqnarray}

where, $\eta$ is a viscosity and $\left\langle \cdot\right\rangle $
represents the Macaulay bracket (ramp function).

From \ref{eq:1_9}-\ref{eq:1_10}, the flow rules for plasticity and
viscoplasticity yield

\begin{eqnarray}
\dot{\epsilon}_{ij}^{p} & = & \dot{\lambda}\;\frac{\partial f}{\partial\sigma_{ij}}\nonumber \\
 & = & \dot{\lambda}\; s_{ij}\label{eq:1_11}
\end{eqnarray}

and

\begin{eqnarray}
\dot{\epsilon}_{ij}^{vp} & = & \frac{\partial\varphi}{\partial s_{ij}}\nonumber \\
 & = & \frac{\partial\varphi}{\partial J_{2}}\frac{\partial J_{2}}{\partial s_{ij}}\nonumber \\
 & = & \frac{1}{2\,\eta}\left\langle 1-\frac{k}{\sqrt{J_{2}}}\right\rangle s_{ij}\label{eq:1_12}
\end{eqnarray}

In \ref{eq:1_11}-\ref{eq:1_12}, $\dot{\epsilon}_{ij}^{p}$, $\dot{\lambda}$,
and $\dot{\epsilon}_{ij}^{vp}$ represent the plastic strain rate,
plastic multiplier, and viscoplastic strain rate, respectively.

Despite such fundamental uses of decomposition in mechanics, the conventional
decomposition method is not directly adopted to the formulation in
FEM and BEM due to the lack of the generalized decomposition method
including decomposition of constitutive or compliance relation. As
we shall see, through the multiplication decomposition that developed
here, one can directly decompose not only stress and strain, but also
constitutive relation and compliance relation. The method is so simple
and general that just a decomposition multiplier is applied to all
cases mentioned here. And as an example, we will also show its applications
to the formulation in FEM and BEM for elastostatics.

\section{Decomposition in tensor forms}

In this Section, we show the multiplying decomposition method for
stress, strain, constitutive relation, and compliance relation in
tensor forms. For canonical example, we also show its application
to linear isotropic elasticity.

\subsection{Decomposition of stress and constitutive relation in tensor form}

Conventional stress decompositions such as \ref{eq:1_4}-\ref{eq:1_5}
can be written in tensor form as

\begin{eqnarray}
s_{ij} & = & \sigma_{ij}-p\delta_{ij}\nonumber \\
 & = & \sigma_{ij}-\frac{1}{3}\sigma_{kk}\delta_{ij}\nonumber \\
 & = & \left(\delta_{ki}\delta_{lj}-\frac{1}{3}\delta_{lk}\delta_{ij}\right)\sigma_{kl}\label{eq:2_1}
\end{eqnarray}

and

\begin{eqnarray}
p & = & \frac{1}{3}\sigma_{kk}\nonumber \\
 & = & \left(\frac{1}{3}\delta_{lk}\right)\sigma_{kl}\label{eq:2_2}
\end{eqnarray}

In \ref{eq:2_1}-\ref{eq:2_2}, if we let 

\begin{equation}
M_{ijkl}^{d}=\delta_{ki}\delta_{lj}-\frac{1}{3}\delta_{lk}\delta_{ij}\label{eq:2_3}
\end{equation}

and

\begin{equation}
M_{ijkl}^{v}=\frac{1}{3}\delta_{lk}\delta_{ij}\label{eq:2_4}
\end{equation}

then, \ref{eq:2_1} and \ref{eq:2_2} can be written as:

\begin{equation}
s_{ij}=M_{ijkl}^{d}\sigma_{kl}\label{eq:2_5}
\end{equation}

and

\begin{equation}
p{}_{ij}=M_{ijkl}^{v}\sigma_{kl}\label{eq:2_6}
\end{equation}

In \ref{eq:2_6}, $p{}_{ij}$ represents

\begin{equation}
p_{ij}=p\delta_{ij}\label{eq:2_7}
\end{equation}

Such decompositions as \ref{eq:2_5} and \ref{eq:2_6} can be easily
checked through

\begin{eqnarray}
\sigma_{ij} & = & s_{ij}+p_{ij}\nonumber \\
 & = & M_{ijkl}^{d}\sigma_{kl}+M_{ijkl}^{v}\sigma_{kl}\nonumber \\
 & = & \left(\delta_{ki}\delta_{lj}-\frac{1}{3}\delta_{lk}\delta_{ij}\right)\sigma_{kl}+\left(\frac{1}{3}\delta_{lk}\delta_{ij}\right)\sigma_{kl}\nonumber \\
 & = & \left(\delta_{ki}\delta_{lj}\right)\sigma_{kl}\nonumber \\
 & = & \sigma_{ij}\label{eq:2_8}
\end{eqnarray}

and, by them, we can also decompose the constitutive relation $C_{ijkl}$
in 

\begin{equation}
\sigma_{ij}=C_{ijkl}\epsilon_{kl}\label{eq:2_9}
\end{equation}

That is, by \ref{eq:2_9} and \ref{eq:2_5}, the deviatoric stress
$s_{ij}$ can be written by

\begin{eqnarray}
s_{ij} & = & M_{ijkl}^{d}\sigma_{kl}\nonumber \\
 & = & M_{ijkl}^{d}\left(C_{klmn}\epsilon_{mn}\right)\nonumber \\
 & = & \left(M_{ijkl}^{d}C_{klmn}\right)\epsilon_{mn}\label{eq:2_10}
\end{eqnarray}

and, by \ref{eq:2_9} and \ref{eq:2_6}, the pressure $p_{ij}$ can
be written by

\begin{eqnarray}
p_{ij} & = & M_{ijkl}^{v}\sigma_{kl}\nonumber \\
 & = & M_{ijkl}^{v}\left(C_{klmn}\epsilon_{mn}\right)\nonumber \\
 & = & \left(M_{ijkl}^{v}C_{klmn}\right)\epsilon_{mn}\label{eq:2_11}
\end{eqnarray}

Then, noting $M_{ijkl}^{d}C_{klmn}$ and $M_{ijkl}^{v}C_{klmn}$ as 

\begin{equation}
C_{ijmn}^{d}=M_{ijkl}^{d}C_{klmn}\label{eq:2_12}
\end{equation}

and

\begin{equation}
C_{ijmn}^{v}=M_{ijkl}^{v}C_{klmn}\label{eq:2_13}
\end{equation}

yields stress decompositions

\begin{eqnarray}
s_{ij} & = & C_{ijmn}^{d}\epsilon_{mn}\label{eq:2_14}\\
p_{ij} & = & C_{ijmn}^{v}\epsilon_{mn}\label{eq:2_15}
\end{eqnarray}

in terms of decomposed constitutive tensors $C_{ijmn}^{d}$ and $C_{ijmn}^{v}$. 

Such constitutive decompositions as \ref{eq:2_12} and \ref{eq:2_13}
are also checked through

\begin{eqnarray}
C_{ijmn}^{d}+C_{ijmn}^{v} & = & \left(M_{ijkl}^{d}+M_{ijkl}^{v}\right)C_{klmn}\nonumber \\
 & = & \delta_{ki}\delta_{lj}C_{klmn}\nonumber \\
 & = & C_{ijmn}\label{eq:2_16}
\end{eqnarray}

\subsection{Decomposition of strain and compliance relation in tensor form}

Conventional strain decompositions are given by

\begin{equation}
\epsilon_{ij}^{'}=\epsilon_{ij}-\frac{1}{3}\epsilon_{kk}\delta_{ij}\label{eq:2_17}
\end{equation}

\begin{equation}
\epsilon_{ij}^{M}=\frac{1}{3}\epsilon_{kk}\delta_{ij}\label{eq:2_18}
\end{equation}

By following similar decomposition procedures to stress, the multiplying
decomposition can define strain decompositions as

\begin{eqnarray}
\epsilon_{ij}^{'} & = & M_{ijkl}^{v}\epsilon_{kl}\label{eq:2_19}\\
\epsilon_{ij}^{M} & = & M_{ijkl}^{d}\epsilon_{kl}\label{eq:2_20}
\end{eqnarray}

and, by them, we also can decompose the compliance relation $D_{ijkl}$
in

\begin{equation}
\epsilon_{ij}=D_{ijkl}\sigma_{kl}\label{eq:2_21}
\end{equation}

into

\begin{equation}
D_{ijmn}^{d}=M_{ijkl}^{d}D_{klmn}\label{eq:2_22}
\end{equation}

and

\begin{equation}
D_{ijmn}^{v}=M_{ijkl}^{v}D_{klmn}\label{eq:2_23}
\end{equation}

In \ref{eq:2_20}, $\epsilon_{ij}^{M}$ represents

\begin{eqnarray}
\epsilon_{ij}^{M} & = & \epsilon_{M}\delta_{ij}\label{eq:2_24}
\end{eqnarray}

and equations \ref{eq:2_19}-\ref{eq:2_20} can also be written by

\begin{eqnarray}
\epsilon_{ij}^{'} & = & D_{ijmn}^{d}\sigma_{mn}\label{eq:2_25}\\
\epsilon_{ij}^{M} & = & D_{ijmn}^{v}\sigma_{mn}\label{eq:2_26}
\end{eqnarray}

in terms of the decomposed compliance tensors $D_{ijmn}^{d}$ and
$D_{ijmn}^{v}$.

Such compliance decompositions as \ref{eq:2_22} and \ref{eq:2_23}
are also checked through

\begin{eqnarray}
D_{ijmn}^{d}+D_{ijmn}^{v} & = & \left(M_{ijkl}^{d}+M_{ijkl}^{v}\right)D_{klmn}\nonumber \\
 & = & \delta_{ki}\delta_{lj}D_{klmn}\nonumber \\
 & = & D_{ijmn}\label{eq:2_27}
\end{eqnarray}

\subsection{Example for linear isotropic elasticity}

For linear isotropic elasticity, the constitutive relation $C_{ijkl}$
is given by

\begin{equation}
C_{ijkl}=\lambda\delta_{kl}\delta_{ij}+2\mu\delta_{ki}\delta_{lj}\label{eq:2_28}
\end{equation}

By the suggested decomposition method, the constitutive decomposition
tensors $C_{ijmn}^{d}$ and $C_{ijmn}^{v}$ are expressed as

\begin{eqnarray}
C_{ijmn}^{d} & = & M_{ijkl}^{d}C_{klmn}\nonumber \\
 & = & \left(\delta_{ki}\delta_{lj}-\frac{1}{3}\delta_{lk}\delta_{ij}\right)\left(\lambda\delta_{mn}\delta_{kl}+2\mu\delta_{mk}\delta_{nl}\right)\nonumber \\
 & = & \frac{2}{3}\mu\left(3\delta_{mi}\delta_{nj}-\delta_{ij}\delta_{mn}\right)\label{eq:2_29}
\end{eqnarray}

and

\begin{eqnarray}
C_{ijmn}^{v} & = & M_{ijkl}^{v}C_{klmn}\nonumber \\
 & = & \left(\frac{1}{3}\delta_{lk}\delta_{ij}\right)\left(\lambda\delta_{mn}\delta_{kl}+2\mu\delta_{mk}\delta_{nl}\right)\nonumber \\
 & = & \frac{1}{3}(3\lambda+2\mu)\delta_{ij}\delta_{mn}\nonumber \\
 & = & \left(\lambda+\frac{2}{3}\mu\right)\delta_{ij}\delta_{mn}\nonumber \\
 & = & K\delta_{ij}\delta_{mn}\label{eq:2_30}
\end{eqnarray}

where $\mu$ and $K$ are shear modulus and bulk modulus, respectively. 

Equations \ref{eq:2_29}-\ref{eq:2_30} can also be checked through
the conventional decomposition method. That is, from the elastic constitutive
relation \ref{eq:2_28}, we have

\begin{eqnarray}
\sigma_{ij} & = & C_{ijmn}\epsilon_{mn}\nonumber \\
 & = & \left(\lambda\delta_{mn}\delta_{ij}+2\mu\delta_{mi}\delta_{nj}\right)\epsilon_{mn}\label{eq:2_31}
\end{eqnarray}

Letting $ij\rightarrow kk$ in \ref{eq:2_31} yields 

\begin{eqnarray}
\sigma_{kk} & = & (3\lambda+2\mu)\epsilon_{kk}\nonumber \\
 & = & 3K\epsilon_{kk}\label{eq:2_32}
\end{eqnarray}

Then, $p{}_{ij}$ that given by \ref{eq:2_7} is expressed as 

\begin{eqnarray}
p{}_{ij} & = & \frac{\sigma_{kk}}{3}\delta_{ij}\nonumber \\
 & = & K\epsilon_{kk}\delta_{ij}\label{eq:2_33}
\end{eqnarray}

Also, $s_{ij}$ given by \ref{eq:2_1} is expressed as

\begin{eqnarray}
s_{ij} & = & \sigma_{ij}-p_{ij}\nonumber \\
 & = & \sigma_{ij}-\frac{\sigma_{kk}}{3}\delta_{ij}\nonumber \\
 & = & \left(\lambda\delta_{kl}\delta_{ij}+2\mu\delta_{ki}\delta_{lj}\right)\epsilon_{kl}-\frac{1}{3}\left((3\lambda+2\mu)\epsilon_{kk}\right)\delta_{ij}\nonumber \\
 & = & \frac{2}{3}\mu\left(3\epsilon_{ij}-\delta_{ij}\epsilon_{kk}\right)\label{eq:2_34}
\end{eqnarray}

Thus, from \ref{eq:2_33} and \ref{eq:2_34}, we can verify the constitutive
relations such as

\begin{equation}
p{}_{ij}=C_{ijmn}^{v}\epsilon_{mn}\label{eq:2_35}
\end{equation}
 and

\begin{equation}
s_{ij}=C_{ijmn}^{d}\epsilon_{mn}\label{eq:2_36}
\end{equation}

Similarly, the compliance tensor $D_{ijkl}$ for linear isotropic
elasticity, which is given by

\begin{equation}
D_{ijkl}=\frac{1}{E}\left((1+\nu)\delta_{ki}\delta_{lj}-\nu\delta_{kl}\delta_{ij}\right)\label{eq:2_37}
\end{equation}

can be decomposed into 

\begin{eqnarray}
D_{ijmn}^{d} & = & M_{ijkl}^{d}D_{klmn}\nonumber \\
 & = & \frac{1}{2\mu}\left(\delta_{im}\delta_{jn}-\frac{1}{3}\delta_{ij}\delta_{mn}\right)\label{eq:2_38}
\end{eqnarray}

and

\begin{eqnarray}
D_{ijmn}^{v} & = & M_{ijkl}^{v}D_{klmn}\nonumber \\
 & = & \frac{1}{3K}\delta_{ij}\delta_{mn}\label{eq:2_39}
\end{eqnarray}

Such equations as \ref{eq:2_38} and \ref{eq:2_39} can also be checked
through the conventional decomposition method.

\section{Decomposition in matrix form}

In this Section, we show the multiplying decomposition method for
stress, strain, constitutive relation, and compliance relation in
matrix forms.

\subsection{Decomposition of stress and constitutive relation in matrix form}

With Voigt notation, the decomposition multiplier \ref{eq:2_3} can
be expressed in a matrix form as

\begin{equation}
\mathbf{\boldsymbol{\mathbf{M^{d}}}}=\frac{1}{3}\begin{bmatrix}2 & -1 & -1 & 0 & 0 & 0\\
-1 & 2 & -1 & 0 & 0 & 0\\
-1 & -1 & 2 & 0 & 0 & 0\\
0 & 0 & 0 & 3 & 0 & 0\\
0 & 0 & 0 & 0 & 3 & 0\\
0 & 0 & 0 & 0 & 0 & 3
\end{bmatrix}\label{eq:3_1}
\end{equation}

Also, the decomposition multiplier \ref{eq:2_4} can be expressed
in a matrix form as

\begin{equation}
\mathbf{\boldsymbol{\mathbf{M^{v}}}}=\frac{1}{3}\begin{bmatrix}1 & 1 & 1 & 0 & 0 & 0\\
1 & 1 & 1 & 0 & 0 & 0\\
1 & 1 & 1 & 0 & 0 & 0\\
0 & 0 & 0 & 0 & 0 & 0\\
0 & 0 & 0 & 0 & 0 & 0\\
0 & 0 & 0 & 0 & 0 & 0
\end{bmatrix}\label{eq:3_2}
\end{equation}

Then, we can express the deviatoric stress in \ref{eq:2_5} and the
pressure in \ref{eq:2_6} in a vector form $\mathbf{s}$ and $\mathbf{p}$
as

\begin{equation}
\mathbf{\boldsymbol{s=M^{d}\sigma}}\label{eq:3_3}
\end{equation}

and

\begin{equation}
\boldsymbol{\mathbf{p=M^{v}\sigma}}\label{eq:3_4}
\end{equation}

where, the Cauchy stress $\mathrm{\boldsymbol{\sigma}}$ can also
be expressed in terms of the constitutive relation matrix $\mathbf{C}$
as

\begin{equation}
\mathrm{\mathbf{\boldsymbol{\sigma=C\epsilon}}}\label{eq:3_5}
\end{equation}

Substituting \ref{eq:3_5} into \ref{eq:3_3} and \ref{eq:3_4} yields 

\begin{eqnarray}
\boldsymbol{\mathbf{s}} & = & \boldsymbol{\mathbf{C^{d}}\;\mathbf{\epsilon}}\label{eq:3_6}\\
\mathbf{\boldsymbol{p}} & = & \boldsymbol{\mathbf{C^{v}}\;\mathbf{\epsilon}}\label{eq:3_7}
\end{eqnarray}

where, the constitutive relation $\mathbf{C}$ is decomposed into
the matrices $\mathbf{C^{d}}$ and $\mathbf{C^{v}}$:

\begin{eqnarray}
\boldsymbol{\mathbf{C^{d}}} & = & \boldsymbol{\mathbf{M^{d}}\;\mathbf{C}}\label{eq:3_8}\\
\mathbf{\boldsymbol{C^{v}}} & = & \boldsymbol{\mathbf{M^{v}}\;\mathbf{C}}\label{eq:3_9}
\end{eqnarray}

Matrix decompositions \ref{eq:3_3}-\ref{eq:3_5} can be checked through

\begin{eqnarray}
\boldsymbol{\mathbf{\sigma}} & = & \boldsymbol{\mathbf{s}+\boldsymbol{p}}\nonumber \\
 & = & \mathbf{\boldsymbol{M^{d}\sigma}+\boldsymbol{M^{v}\sigma}}\nonumber \\
 & = & \left(\boldsymbol{\mathbf{M^{d}}+\mathbf{M^{v}}}\right)\boldsymbol{\sigma}\nonumber \\
 & = & \boldsymbol{\mathbf{I\sigma}}\nonumber \\
 & = & \mathbf{\boldsymbol{\sigma}}\label{eq:3_10}
\end{eqnarray}

where, $\mathbf{I}$ in \ref{eq:3_10} is the identity matrix of size
6.

The matrix decompositions \ref{eq:3_6}-\ref{eq:3_9} can be also
checked through

\begin{eqnarray}
\mathbf{\boldsymbol{\sigma}} & = & \boldsymbol{\mathbf{s}+\mathbf{p}}\nonumber \\
 & = & \boldsymbol{\mathbf{C^{d}}\;\mathbf{\epsilon}}+\mathbf{\mathbf{C^{v}}}\;\mathbf{\boldsymbol{\epsilon}}\nonumber \\
 & = & \boldsymbol{\left(\mathbf{M^{d}}\;\mathbf{C}+\mathbf{M^{v}}\;\mathbf{C}\right)\;\mathbf{\epsilon}}\nonumber \\
 & = & \left(\mathbf{M^{d}}+\mathbf{M^{v}}\right)\boldsymbol{\;\mathbf{C}\;\mathbf{\epsilon}}\nonumber \\
 & = & \boldsymbol{\mathbf{C}\;\mathbf{\epsilon}}\label{eq:3_11}
\end{eqnarray}

\subsection{Decomposition of strain and compliance relation in matrix form}

Similarly, we have the matrix decompositions for strain and compliance
relation as

\begin{eqnarray}
\boldsymbol{\mathbf{\epsilon^{d}}} & = & \boldsymbol{\mathbf{M^{d}}\epsilon}\label{eq:3_12}\\
\boldsymbol{\epsilon^{v}} & = & \boldsymbol{\mathbf{M^{v}}\epsilon}\label{eq:3_13}\\
\boldsymbol{\mathbf{\epsilon^{d}}} & = & \mathbf{\boldsymbol{D^{d}\sigma}}\label{eq:3_14}\\
\boldsymbol{\mathbf{\epsilon^{v}}} & = & \mathbf{\boldsymbol{D^{v}\sigma}}\label{eq:3_15}
\end{eqnarray}

and

\begin{eqnarray}
\boldsymbol{\mathbf{D^{d}}} & = & \mathbf{\boldsymbol{M^{d}\; D}}\label{eq:3_16}\\
\boldsymbol{\mathbf{D^{v}}} & = & \boldsymbol{\mathbf{M^{v}\; D}}\label{eq:3_17}
\end{eqnarray}

In \ref{eq:3_12}-\ref{eq:3_17}, $\boldsymbol{\mathbf{\epsilon^{d}}}$,$\boldsymbol{\mathbf{\epsilon^{v}}}$,
$\mathbf{D^{d}}$, and $\mathbf{D^{v}}$ represent the deviatoric
strain, the volumetric strain, the deviatoric compliance relation,
and the volumetric compliance relation in a vector and matrix form,
respectively.

\section{Properties of multiplying decomposition method}

So far, we have shown the multiplying decomposition method for stress,
strain, constitutive relation, and compliance relation in both tensor
and matrix forms. The method just takes the multiplication decomposers
$M_{ijkl}^{d}$ (or $\mathbf{M^{d}}$) and $M_{ijkl}^{v}$ (or $\mathbf{M^{v}}$)
and we will see their properties in this Section. As we shall see,
the multiplication decomposition is also valid to decompose strain
energy density and compatible with physical meaning.

\subsection{Decomposition of strain energy}

With the multiplying decomposition method, strain energy density $u$
can be written as

\begin{eqnarray}
u & = & \frac{1}{2}\; C_{ijkl}\;\epsilon_{ij}\;\epsilon_{kl}\nonumber \\
 & = & \frac{1}{2}\;\left(C_{ijkl}^{d}+C_{ijkl}^{v}\right)\epsilon_{ij}\;\epsilon_{kl}\nonumber \\
 & = & \frac{1}{2}\;\left(s_{kl}+p_{kl}\right)\;\left(\epsilon_{kl}^{'}+\epsilon_{kl}^{M}\right)\nonumber \\
 & = & \frac{1}{2}\; s_{kl}\;\epsilon_{kl}^{'}+\frac{1}{2}\; p_{kl}\;\epsilon_{kl}^{M}\label{eq:4_1}
\end{eqnarray}

While deriving \ref{eq:4_1}, we use the relations

\begin{eqnarray}
s_{kl}\;\epsilon_{kl}^{M} & = & 0\label{eq:4_2}\\
p_{kl}\;\epsilon_{kl}^{'} & = & 0\label{eq:4_3}
\end{eqnarray}

since we have

\begin{eqnarray}
s_{kl}\;\epsilon_{kl}^{M} & = & s_{kl}\;\epsilon_{M}\delta_{kl}\nonumber \\
 & = & \epsilon_{M}\; s_{kk}\nonumber \\
 & = & 0\label{eq:4_4}
\end{eqnarray}

and

\begin{eqnarray}
p_{kl}\;\epsilon_{kl}^{'} & = & p\delta_{kl}\;\left(\epsilon_{kl}-\epsilon_{M}\delta_{kl}\right)\nonumber \\
 & = & p\left(\epsilon_{kk}-\epsilon_{M}\delta_{kk}\right)\nonumber \\
 & = & 0\label{eq:4_5}
\end{eqnarray}

In \ref{eq:4_1}, the multiplying decomposition method also decomposes
the strain energy density $u$ into deviatoric strain energy density
$\left(\frac{1}{2}\; s_{kl}\;\epsilon_{kl}^{'}\right)$ and volumetric
strain energy density $\left(\frac{1}{2}\; p_{kl}\;\epsilon_{kl}^{M}\right)$.
Such strain energy density decomposition can also be derived from
$u=\frac{1}{2}\; D{}_{ijkl}\;\sigma_{ij}\;\sigma_{kl}$.

\subsection{Compatibility with physical meaning}

With the multiplying decomposition method, one may consider this can
be also used for 
\begin{itemize}
\item the evaluation of total stress from either deviatoric stress (strain)
or pressure (volumetric strain)
\item the evaluation of total strain from either deviatoric strain (stress)
or volumetric strain (pressure)
\end{itemize}
However, this is not possible because both the decomposition matrices
$\boldsymbol{\mathbf{M^{d}}}$ and $\mathbf{\boldsymbol{M^{v}}}$
are singular. The eigenvalues of $\boldsymbol{\mathbf{M^{d}}}$ are
$\lambda_{1}=0$ and $\lambda_{2}=1$ with multiplicity 5 while the
eigenvalues of $\mathbf{\boldsymbol{M^{v}}}$are $\lambda_{1}=0$
with multiplicity 5 and $\lambda_{2}=1$. In fact, the decomposition
matrices $\boldsymbol{\mathbf{M^{d}}}$ and $\boldsymbol{\mathbf{M^{v}}}$
are projection matrices. And having the eigenvalue of multiplicity
5 in $\boldsymbol{\mathbf{M^{d}}}$ and $\boldsymbol{\mathbf{M^{v}}}$
are natural since $\boldsymbol{\mathbf{M^{d}}}$ leaves only five
deviatoric stress (or deviatoric strain) components and $\mathbf{\boldsymbol{M^{v}}}$
leaves only one pressure (or volumetric strain) component when multiplied
to stress (or strain). Thus, we cannot have $\left[\boldsymbol{\mathbf{M^{d}}}\right]^{-1}$
and $\left[\boldsymbol{\mathbf{M^{v}}}\right]^{-1}$ in \ref{eq:3_3}-\ref{eq:3_4}
and \ref{eq:3_12}-\ref{eq:3_13}. Consequently, we also cannot have
$\left[\boldsymbol{\mathbf{C^{d}}}\right]^{-1}$,$\left[\mathbf{\boldsymbol{C^{v}}}\right]^{-1}$,
$\left[\boldsymbol{\mathbf{D^{d}}}\right]^{-1}$, and $\left[\mathbf{\boldsymbol{D^{v}}}\right]^{-1}$in
\ref{eq:3_8}-\ref{eq:3_9} and \ref{eq:3_16}-\ref{eq:3_17}. 

In physical viewpoint, the properties of multiplying decomposition
can be interpreted as:
\begin{itemize}
\item Total stress results in both deviatoric and volumetric strain but
not vice versa
\item Total strain results in both deviatoric stress and pressure but not
vice versa
\end{itemize}

\section{Applications}

The multiplying decomposition method can be directly used to formulate
both FEM and BEM. In this Section, we show its canonical application
to elastostatics.

\subsection{Finite element formulation}

In FEM for elastostatics (see \citet{strang1973analysis,NEW_Cook,NEW_Bathe,NEW_Hughes,braess2001finite}),
the element stiffness matrix $\mathbf{\boldsymbol{K_{e}}}$ is determined
by 

\begin{equation}
\boldsymbol{\mathbf{K_{e}}}=\int_{\Omega_{e}}\;\mathbf{\boldsymbol{B^{T}\; C\; B}}\; d\Omega\label{eq:5_1}
\end{equation}

where, $\boldsymbol{\mathbf{B}}$,$\mathbf{\boldsymbol{C}}$, and
$\Omega_{e}$ are the strain-displacement matrix, the constitutive
relation matrix, and the domain of the element, respectively.

After assembling all the stiffness matrices of each element and accounting
for boundary conditions to describe the given structure, the global
system of equations is given by

\begin{equation}
\boldsymbol{\mathbf{K\; u=f}}\label{eq:5_2}
\end{equation}

In \ref{eq:5_2}, $\mathbf{\boldsymbol{K}}$,$\boldsymbol{u}$ and
$\mathbf{\boldsymbol{f}}$ represent the global stiffness matrix,
nodal displacements and nodal forces in a vector form, respectively. 

With the adoption of the multiplying decomposition, we can reformulate
the element stiffness matrix $\boldsymbol{\mathbf{K_{e}}}$ in \ref{eq:5_1}
as

\begin{eqnarray}
\boldsymbol{\mathbf{K_{e}}} & = & \int_{\Omega_{e}}\;\mathbf{\boldsymbol{B^{T}}\left(\mathbf{\boldsymbol{C^{d}}}\mathbf{+}\boldsymbol{C^{v}}\right)\boldsymbol{B}}\; d\Omega\nonumber \\
 & = & \int_{\Omega_{e}}\;\mathbf{\boldsymbol{B^{T}}}\boldsymbol{\mathbf{C^{d}}}\boldsymbol{\mathbf{B}}\; d\Omega+\int_{\Omega_{e}}\;\boldsymbol{\mathbf{B^{T}}\mathbf{C^{v}}}\mathbf{\boldsymbol{B}}\; d\Omega\label{eq:K_e}\\
 & = & \boldsymbol{\mathbf{K_{e}^{d}}}+\boldsymbol{\mathbf{K_{e}^{v}}}\nonumber 
\end{eqnarray}

where $\boldsymbol{\mathbf{K_{e}^{d}}}$ and $\boldsymbol{\mathbf{K_{e}^{v}}}$
are decomposed element deviatoric stiffness matrix and element volumetric
stiffness matrix.

Also, the global system of equations can be written as

\begin{equation}
\left(\mathbf{\boldsymbol{K^{d}}}+\mathbf{\boldsymbol{K^{v}}}\right)\boldsymbol{\mathbf{\; u=f}}\label{eq:5_4}
\end{equation}

\begin{equation}
\mathbf{\boldsymbol{\boldsymbol{K}}}=\mathbf{\boldsymbol{K^{d}}}+\mathbf{\boldsymbol{K^{v}}}\label{eq:K=00003DK^d+K^v}
\end{equation}

As in \ref{eq:K=00003DK^d+K^v}, with the multiplying decomposition,
the global stiffness matrix $\mathbf{\boldsymbol{K}}$ is decomposed
into the global deviatoric stiffness matrix $\mathbf{\boldsymbol{K^{d}}}$
and the global volumetric stiffness matrix $\mathbf{\boldsymbol{K^{v}}}$,
where $\boldsymbol{\mathbf{K^{d}}}$ results from assembling all the
$\boldsymbol{\mathbf{K_{e}^{d}}}$ and $\mathbf{\boldsymbol{\boldsymbol{K^{v}}}}$
results from assembling all the $\boldsymbol{\mathbf{K_{e}^{v}}}$.

Separate evaluation of $\boldsymbol{\mathbf{K_{e}^{d}}}$ and $\boldsymbol{\mathbf{K_{e}^{v}}}$
by the multiplying decomposition allows ``selective integration technique''
in a direct way and some useful applications are listed below. 
\begin{itemize}
\item Using the reduced integration for volumetric part and full integration
for deviatoric part: in \emph{(nearly) incompressible cases,} \emph{selectively
reduced-integration technique (SRI)} is used for 1st order linear
isoparametric elements (4-node elements in two dimensions and 8-node
elements in three dimensions), to prevent mesh locking and get accurate
solution for (nearly) incompressible cases (see \citet{malkus1978mixed,hughes2005generalization,doll2000volumetric,NEWliu1994multiple,NEWliu1998multiple}). 
\item Using the reduced integration for deviatoric part and full integration
for volumetric part: \emph{shear locking} can be also prevented by
\emph{SRI}. (See \citet{NEW_Bathe,NEW_Cook,braess2001finite,NEW_Hughes})
\item Formulation of plasticity: when von Mises \emph{plasticity} occurs
(see \ref{eq:1_9}, \ref{eq:1_11}), \emph{$\boldsymbol{K^{d}}$ is
evaluated iteratively} while $\boldsymbol{K^{v}}$ remains constant
since $\epsilon_{kk}=0$ (See \citet{NEWdunne2005introduction,NEW_lubliner1990plasticity}).
With the adoption of the multiplying decomposition, we can separately
evaluate $\boldsymbol{K^{d}}$ without evaluating $\boldsymbol{K^{v}}$
during plastic evolution (computation can be reduced). 
\end{itemize}

\subsection{Boundary element formulation}

In BEM, an integral equation for interior stress in elasticity can
be written as (see \citet{banerjee_boundary_1994,beer_boundary_2010,ang_beginners_2007,wrobel_boundary_2002})

\begin{equation}
\sigma_{ij}(\boldsymbol{\xi})=\int_{S}\left(G_{kij}^{\sigma}(\boldsymbol{x},\boldsymbol{\xi})t_{k}(\boldsymbol{x})-F_{kij}^{\sigma}(\boldsymbol{x},\boldsymbol{\xi})u_{k}(\boldsymbol{x})\right)\mathrm{d}S(\boldsymbol{x})
\end{equation}

where $G_{kij}^{\sigma}$ and $F_{kij}^{\sigma}$ are traction and
displacement kernel function for stress respectively, and $S$ means
the boundary of a given problem domain.

With the multiplication decomposition, deviatoric stress can be derived
as

\begin{eqnarray}
s_{mn}(\boldsymbol{\xi}) & = & M_{ijmn}^{d}\sigma_{ij}(\boldsymbol{\xi})\nonumber \\
 & = & M_{ijmn}^{d}\int_{S}\left(G_{kij}^{\sigma}(\boldsymbol{x},\boldsymbol{\xi})t_{k}(\boldsymbol{x})-F_{kij}^{\sigma}(\boldsymbol{x},\boldsymbol{\xi})u_{k}(\boldsymbol{x})\right)\mathrm{d}S(\boldsymbol{x})\label{eq:s_ij BEM}
\end{eqnarray}

Similarly, pressure can be derived as

\begin{eqnarray}
p_{mn}(\boldsymbol{\xi}) & = & M_{ijmn}^{v}\int_{S}\left(G_{kij}^{\sigma}(\boldsymbol{x},\boldsymbol{\xi})t_{k}(\boldsymbol{x})-F_{kij}^{\sigma}(\boldsymbol{x},\boldsymbol{\xi})u_{k}(\boldsymbol{x})\right)\mathrm{d}S(\boldsymbol{x})\label{eq:p_ij BEM}
\end{eqnarray}

Similarly, decomposed strains are as follows

\begin{equation}
\epsilon_{mn}^{'}(\boldsymbol{\xi})=M_{ijmn}^{d}\int_{S}\left(G_{kij}^{\epsilon}(\boldsymbol{x},\boldsymbol{\xi})t_{k}(\boldsymbol{x})-F_{kij}^{\epsilon}(\boldsymbol{x},\boldsymbol{\xi})u_{k}(\boldsymbol{x})\right)\mathrm{d}S(\boldsymbol{x})\label{eq:epsilon'_ij BEM}
\end{equation}

\begin{equation}
\epsilon_{mn}^{M}(\boldsymbol{\xi})=M_{ijmn}^{v}\int_{S}\left(G_{kij}^{\epsilon}(\boldsymbol{x},\boldsymbol{\xi})t_{k}(\boldsymbol{x})-F_{kij}^{\epsilon}(\boldsymbol{x},\boldsymbol{\xi})u_{k}(\boldsymbol{x})\right)\mathrm{d}S(\boldsymbol{x})\label{eq:epsilon^M_ij BEM}
\end{equation}

where $G_{kij}^{\epsilon}$ and $F_{kij}^{\epsilon}$ are traction
and displacement kernel function for strain respectively.

\section{Conclusions}

A simple, clear, and widely applicable way to decompose stress/strain
and constitutive/compliance relations is suggested in both tensor
and matrix forms: multiplication decomposition. The method is also
applicable to decompose strain energy density along with proper physical
meaning. 

We consider here the application of multiplying decomposition to elastostatics
in FEM and BEM formulation, which illustrates the elegance of this
approach. Clearly, however, the multiplying decomposition is quite
general and can be applied readily to elastoplasticity, viscoplasticity,
fluid mechanics and more broadly throughout mechanics. In addition,
we anticipate that the multiplying decomposition method developed
here will provide an interesting foundation for the development of
novel analytic and computational methods.

\bibliographystyle{elsart-harv}
\bibliography{BEM,Bib}

\end{document}